\documentclass{article}

\usepackage{times}
\usepackage{graphicx} 
\usepackage{subfigure} 

\usepackage{natbib}

\usepackage{algorithm}
\usepackage{algorithmic}

\usepackage{hyperref}

\usepackage[accepted]{icml2017}

\usepackage{amsmath,amssymb}

\begin{document} 

\twocolumn[
\icmltitle{Multichannel End-to-end Speech Recognition
}


\icmlauthor{Tsubasa Ochiai}{eup1105@mail4.doshisha.ac.jp}
\icmladdress{Doshisha University}
\icmlauthor{Shinji Watanabe}{watanabe@merl.com}
\icmlauthor{Takaaki Hori}{thori@merl.com}
\icmlauthor{John R. Hershey}{hershey@merl.com}
\icmladdress{Mitsubishi Electric Research Laboratories (MERL)}

\icmlkeywords{}

\vskip 0.3in
]

\begin{abstract} 

The field of speech recognition is in the midst of a  paradigm shift:  end-to-end neural networks are challenging the dominance of hidden Markov models as a  core technology.
Using an attention mechanism in a recurrent encoder-decoder architecture solves the dynamic time alignment problem,  allowing joint end-to-end training of the acoustic and language modeling components.    
In this paper we extend the end-to-end framework to encompass microphone array signal processing for noise suppression and speech enhancement within the acoustic encoding network.
This allows the beamforming components to be optimized jointly within the recognition architecture to improve the end-to-end speech recognition objective.    
Experiments on the noisy speech benchmarks (CHiME-4 and AMI) show that our multichannel end-to-end system outperformed the attention-based baseline with input from a conventional adaptive beamformer.

\end{abstract} 

\section{Introduction}
Existing automatic speech recognition (ASR) systems are based on a complicated hybrid of separate components, including acoustic, phonetic, and language models \cite{jelinek1976continuous}.
Such systems are typically based on deep neural network acoustic models combined with hidden Markov models to represent the language and phonetic context-dependent state and their temporal alignment with the acoustic signal (DNN-HMM) \cite{bourlard1994connectionist,hinton2012deep}.
As a simpler alternative, end-to-end speech recognition paradigm has attracted great research interest  \cite{chorowski2014end,chorowski2015attention,chan2016listen,graves2014towards,miao2015eesen}.  This paradigm simplifies the above hybrid architecture by subsuming it into a single neural network.
Specifically, an attention-based encoder-decoder framework \cite{chorowski2014end} integrates all of those components using a set of recurrent neural networks (RNN), which map from acoustic feature sequences to character label sequences.

However, existing end-to-end frameworks have focused on clean speech, and do not include speech enhancement, which is essential to good performance in noisy environments.
For example, recent industrial applications (e.g., Amazon echo) and benchmark studies \cite{barker2016third,kinoshita2016summary} show that multichannel speech enhancement techniques, using beamforming methods, produce substantial improvements as a pre-processor for conventional hybrid systems, in the presence of strong background noise.
In light of the above trends, this paper extends the existing attention-based encoder-decoder framework by integrating multichannel speech enhancement.  Our proposed  \textit{multichannel end-to-end speech recognition} framework is trained to directly translate from multichannel acoustic signals to text.

A key concept of the multichannel end-to-end framework is to optimize the entire inference procedure, including the beamforming, based on the final ASR objectives, such as word/character error rate (WER/CER).
Traditionally, beamforming techniques such as delay-and-sum and filter-and-sum  
are optimized based on a signal-level loss function, independently of speech recognition task \cite{benesty2008microphone,van1988beamforming}.  Their use in ASR requires ad-hoc modifications such as Wiener post-filtering or distortionless constraints, as well as steering mechanisms determine a look direction to focus the beamformer on the target speech \cite{wolfel2009distant}. 
In contrast, our framework incorporates recently proposed neural beamforming mechanisms as a differentiable component to allow joint optimization of the multichannel speech enhancement within the end-to-end system to improve the ASR objective.    

Recent studies on neural beamformers can be categorized into two types: (1) beamformers with a filter estimation network \cite{xiao2016deep,li2016neural} and (2) beamformers with a mask estimation network \cite{heymann2016neural,erdogan2016improved}.
Both methods obtain an enhanced signal based on the formalization of the conventional filter-and-sum beamformer in the time-frequency domain.
The main difference between them is how the multichannel filters are produced by the neural network.
In the former approach, the multichannel filter coefficients are direct outputs of the network.
In the latter approach, a network first estimates time-frequency masks, which are used to compute expected speech and noise statistics.
Then, using these statistics, the filter coefficients are computed based on the well-known MVDR (minimum variance distortionless response) formalization \cite{capon1969high}.  In both approaches, the estimated filter coefficients are then applied to the multichannel noisy signal to enhance the speech signal. 
Note that the mask estimation approach has the advantage of leveraging well-known techniques, but it requires parallel data composed of aligned clean and noisy speech, which are usually difficult to obtain without data simulation.

Recently, it has been reported that the mask estimation-based approaches \cite{yoshioka2015ntt,heymann2016neural,erdogan2016improved} achieve great performance in noisy speech recognition benchmarks (e.g., CHiME 3 and 4 challenges)\footnote{\citealt{yoshioka2015ntt} uses a clustering technique to perform mask estimation rather than the neural network-based techniques, but it uses the same MVDR formulation for filter estimation.}.
Although this paper proposes to incorporate both mask and filter estimation approaches in an end-to-end framework, motivated by those successes, we focus more on the mask estimation, implementing it along with the MVDR estimation as a differentiable network.  Our MVDR formulation estimates the speech image at the \emph{reference} microphone and includes selection of the reference microphone using an attention mechanism.
By using channel-independent mask estimation along with this reference selection, the model can generalize to different microphone array geometries (number of channels, microphone locations, and ordering), unlike the filter estimation approach.  Finally, because the masks are latent variables in the end-to-end training, we no longer need parallel clean and noisy speech. 

The main advantages of our proposed multichannel end-to-end speech recognition system are:
\begin{enumerate}
\item Overall inference from speech enhancement to recognition is jointly optimized for the ASR objective.
\item The trained system can be used for input signals with arbitrary number and order of channels.
\item Parallel clean and noisy data are not required. We can optimize the speech enhancement component with noisy signals and their transcripts.
\end{enumerate}

\section{Overview of attention-based encoder-decoder networks}
\label{sec:e2e_conventional}
\begin{figure}[t]
\begin{center}
\centerline{\includegraphics[width=\columnwidth]{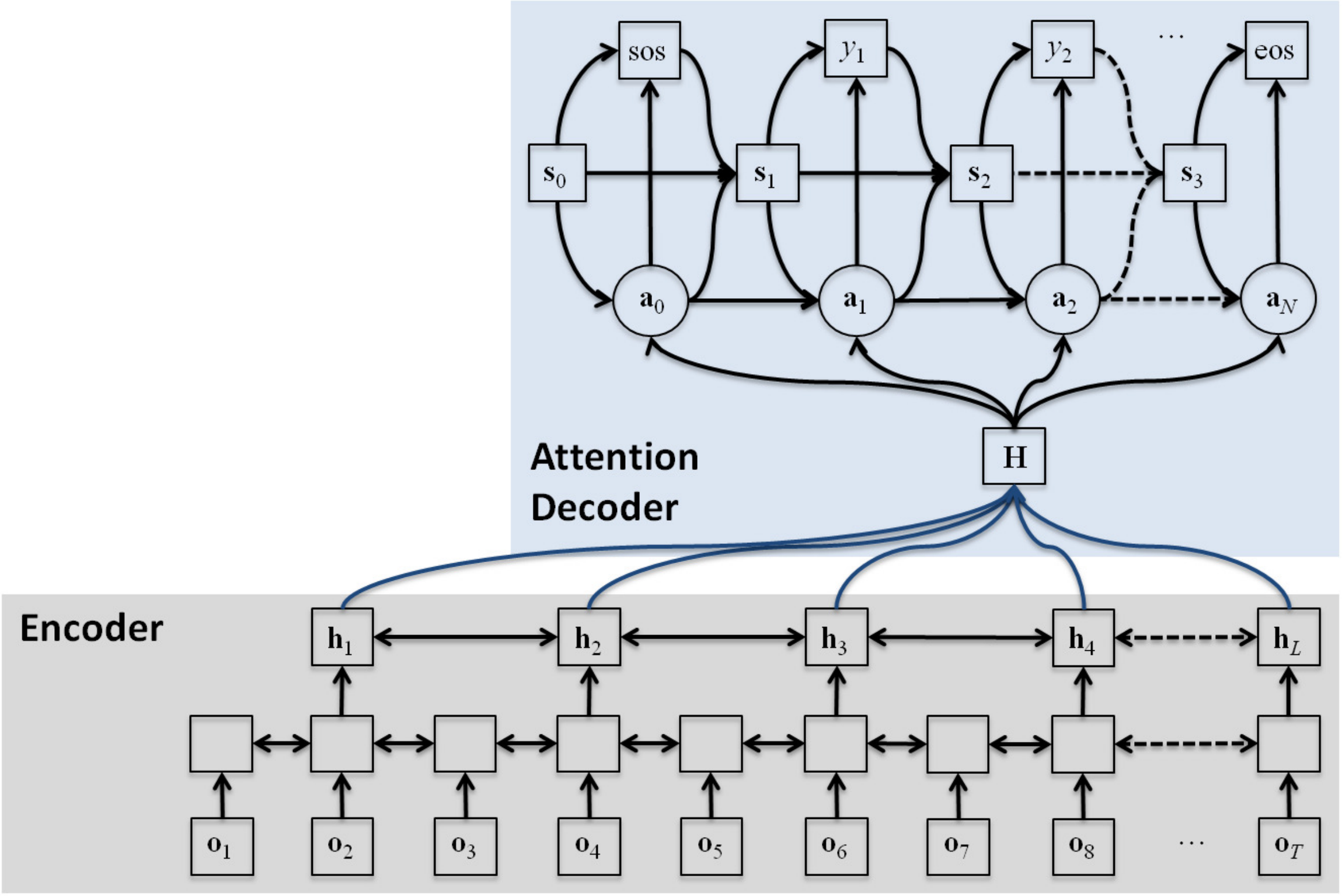}}
\caption{The structure of an attention-based encoder-decoder framework.
The encoder transforms an input sequence $O$ into a high-level feature sequence $H$, and then the decoder generates a character sequence $Y$ through the attention mechanism.}
\label{fig:end-to-end}
\end{center}
\vskip -0.2in
\end{figure} 
This section explains a conventional attention-based encoder-decoder framework, which is used to directly deal with variable length input and output sequences.
The framework consists of two RNNs, called encoder and decoder respectively, and an attention mechanism, which connects the encoder and decoder, as shown in Figure~\ref{fig:end-to-end}.
Given a $T$-length sequence of input features $O = \{\mathbf{o}_{t} \in \mathbb{R} ^{D_\text{O}} | t=1, \cdots, T\}$, the network generates an $N$-length sequence of output labels $Y = \{ y_{n} \in \mathcal{V} | n = 1, \cdots, N \}$, where $\mathbf{o}_{t}$ is a $D_{\text{O}}$-dimensional feature vector (e.g., log Mel filterbank) at input time step $t$, and $y_{n}$ is a label symbol (e.g., character) at output time step $n$ in label set $\mathcal{V}$.

First, given an input sequence $O$, the encoder network transforms it to an $L$-length high-level feature sequence $H = \{\mathbf{h}_{l} \in \mathbb{R}^{D_{\text{H}}} | l = 1, \cdots, L\}$, where $\mathbf{h}_{l}$ is a $D_{\text{H}}$-dimensional state vector at a time step $l$ of encoder's top layer.
In this work, the encoder network is composed of a bidirectional long short-term memory (BLSTM) recurrent network.
To reduce the input sequence length, we apply a subsampling technique \cite{bahdanau2016end} to some layers.
Therefore, $l$ represents the frame index subsampled from $t$ and $L$ is less than $T$.

Next, the attention mechanism integrates all encoder outputs $H$ into a $D_{\text{H}}$-dimensional context vector $\mathbf{c}_{n} \in \mathbb{R}^{D_{\text{H}}}$ based on an $L$-dimensional attention weight vector $\mathbf{a}_{n} \in [0, 1]^{L}$, which represents a soft alignment of encoder outputs at an output time step $n$.
In this work, we adopt a location-based attention mechanism \cite{chorowski2015attention}, and $\mathbf{a}_{n}$ and $\mathbf{c}_{n}$ are formalized as follows:
\begin{align}
\mathbf{f}_{n} &= \mathbf{F} * \mathbf{a}_{n-1}, \label{eq:att_begin} \\
k_{n,l} &= \mathbf{w}^{\mathrm{T}} \mathrm{tanh}(
\mathbf{V}^{\text{S}} \mathbf{s}_{n} + 
\mathbf{V}^{\text{H}} \mathbf{h}_{l} + 
\mathbf{V}^{\text{F}} \mathbf{f}_{n,l} + 
\mathbf{b}), \label{eq:att_mid} \\
a_{n,l} &= \frac{\exp(\alpha k_{n,l})}{\sum_{l=1}^{L} \exp(\alpha k_{n,l})}, \quad \mathbf{c}_{n} = \sum_{l=1}^{L} a_{n,l} \mathbf{h}_{l}, \label{eq:att_end}
\end{align}
where 
$\mathbf{w} \in \mathbb{R}^{1 \times D_\text{W}}$, 
$\mathbf{V}^{\text{H}} \in \mathbb{R}^{D_\text{W} \times D_\text{H}}$,  
$\mathbf{V}^{\text{S}} \in \mathbb{R}^{D_\text{W} \times D_\text{S}}$, 
$\mathbf{V}^{\text{F}} \in \mathbb{R}^{D_\text{W} \times D_\text{F}}$ 
are trainable weight matrices, $\mathbf{b} \in \mathbb{R}^{D_\text{W}}$ is a trainable bias vector, 
$\mathbf{F} \in \mathbb{R}^{D_\text{F} \times 1 \times D_\text{f}}$ is a trainable convolution filter. 
$\mathbf{s}_{n} \in \mathbb{R}^{D_\text{S}}$ is a $D_\text{S}$-dimensional hidden state vector obtained from an upper decoder network at $n$, and $\alpha$ is a sharpening factor \cite{chorowski2015attention}.
$*$ denotes the convolution operation.

Then, the decoder network incrementally updates a hidden state $\mathbf{s}_{n}$ and generates an output label $y_{n}$ as follows:
\begin{align}
\mathbf{s}_{n} &= \mathrm{Update}(\mathbf{s}_{n-1}, \mathbf{c}_{n-1}, y_{n-1}), \label{eq:dec_begin} \\
y_{n} &= \mathrm{Generate}(\mathbf{s}_{n}, \mathbf{c}_{n}), \label{eq:dec_end}
\end{align}
where the $\mathrm{Generate}(\cdot)$ and $\mathrm{Update}(\cdot)$ functions are composed of a feed forward
network and an LSTM-based recurrent network, respectively.

Now, we can summarize these procedures as follows:
%
\begin{align}
P(Y|O) &= \prod_{n} P(y_{n}|O, y_{1:n-1}), \label{eq:prob} \\
H &= \mathrm{Encoder}(O), \label{eq:enc} \\
\mathbf{c}_{n} &= \mathrm{Attention}(\mathbf{a}_{n-1}, \mathbf{s}_{n}, H), \label{eq:att} \\
y_{n} &= \mathrm{Decoder}(\mathbf{c}_{n}, y_{1:n-1}),  \label{eq:dec}
\end{align}
where $\mathrm{Encoder}(\cdot) = \mathrm{BLSTM}(\cdot)$, $\mathrm{Attention}(\cdot)$ corresponds to Eqs.~(\ref{eq:att_begin})-(\ref{eq:att_end}), and $\mathrm{Decoder}(\cdot)$ corresponds to Eqs.~(\ref{eq:dec_begin}) and (\ref{eq:dec_end}).
Here, special tokens for start-of-sentence (sos) and end-of-sentence (eos) are added to the label set $\mathcal{V}$.
The decoder starts the recurrent computation with the (sos) label and continues to generate output labels until the (eos) label is emitted.
Figure~\ref{fig:end-to-end} illustrates such procedures.

Based on the cross-entropy criterion, the loss function is defined using Eq.~(\ref{eq:prob}) as follows:
\begin{align}
\mathcal{L} = -\ln{P(Y^{*}|O)} = -\sum_{n} \ln{P(y_{n}^{*}|O,y_{1:n-1}^{*})}
\label{eq:e2eobj},
\end{align}
where $Y^{*}$ is the ground truth of a whole sequence of output labels and $y_{1:n-1}^{*}$ is the ground truth of its subsequence until an output time step $n-1$.

In this framework, the whole networks including the encoder, attention, and decoder can be optimized to generate the correct label sequence.
This consistent optimization of all relevant procedures is the main motivation of the end-to-end framework.

\section{Neural beamformers}
\label{sec:beamforming_network}
This section explains neural beamformer techniques, which are integrated with the encoder-decoder network in the following section.
This paper uses frequency-domain beamformers rather than time-domain ones, which achieve significant computational complexity reduction in multichannel neural processing \cite{li2016neural,sainath2016reducing}.
In the frequency domain representation, a filter-and-sum beamformer obtains an enhanced signal as follows:

\begin{figure}[t]
\vskip -0.15in
\begin{center}
\subfigure[Beamforming with filter estimation network]{
\includegraphics[width=3.0cm]{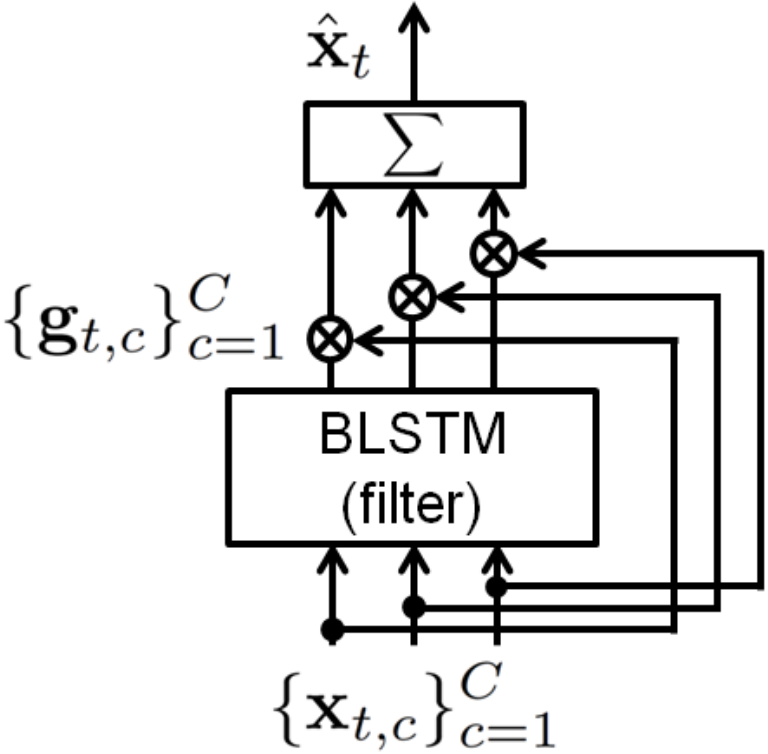}
}
\subfigure[Beamforming with mask estimation network]{
\includegraphics[width=3.0cm]{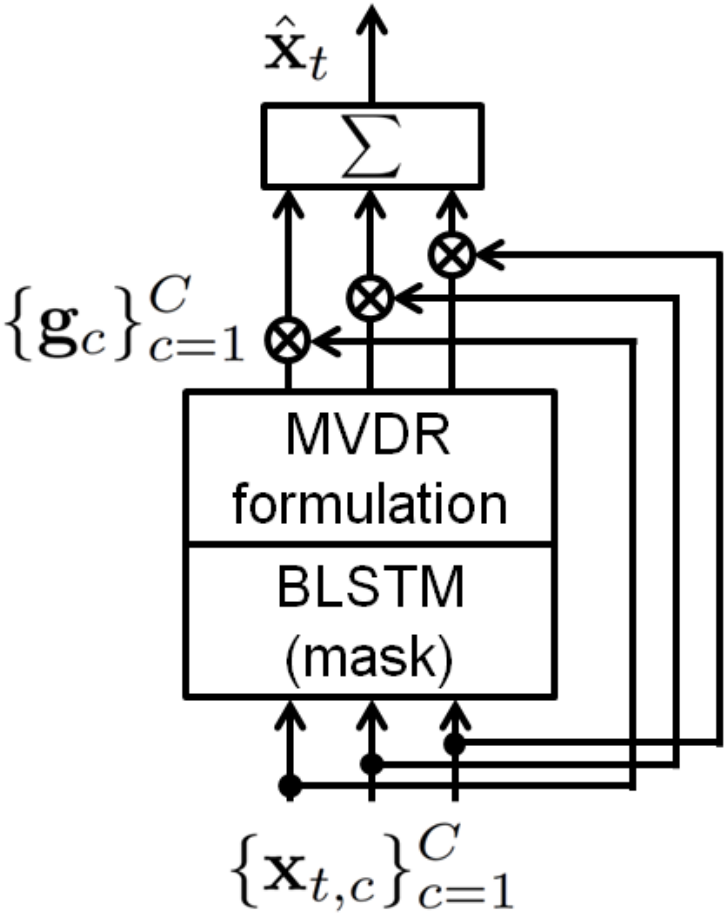}
}
\caption{Structures of neural beamformers. (a) Filter estimation network, which directly estimates the filter coefficients. (b) Mask estimation network, which estimates time-frequency masks, and then get filter coefficients based on the MVDR formalization.}
\label{fig:two_beamformer}
\end{center}
\vskip -0.10in
\end{figure}

\begin{align}
\hat{x}_{t,f} = \sum_{c=1}^{C} g_{t,f,c} x_{t,f,c},
\label{eq:enhance}
\end{align}
where $x_{t,f,c} \in \mathbb{C}$ is an STFT coefficient of $c$-th channel noisy signal at a time-frequency bin $(t, f)$. $g_{t,f,c} \in \mathbb{C}$ is a corresponding beamforming filter coefficient.
$\hat{x}_{t,f} \in \mathbb{C}$ is an enhanced STFT coefficient, and $C$ is the numbers of channels.

In this paper, we adopt two types of neural beamformers, which basically follow Eq.~(\ref{eq:enhance}); 1) filter estimation network and 2) mask estimation network.
Figure~\ref{fig:two_beamformer} illustrates the schematic structure of each approach.
The main difference between them is how to compute the filter coefficient $g_{t,f,c}$.
The following subsections describe each approach.

\subsection{Filter estimation network approach}
\label{sec:filter_net}
The filter estimation network directly estimates a \textit{time-variant} filter coefficients $\{g_{t,f,c}\}_{t=1,f=1,c=1}^{T,F,C}$ as the outputs of the network, which was originally proposed in \cite{li2016neural}.
$F$ is the dimension of STFT features.

This approach uses a single real-valued BLSTM network to predict the real and imaginary parts of the complex-valued filter coefficients at an every time step.
Therefore, we introduce multiple ($2 \times C$) output layers to separately compute the real and imaginary parts of the filter coefficients for each channel.
Then, the network outputs time-variant filter coefficients $\mathbf{g}_{t,c} = \{g_{t,f,c}\}_{f=1}^{F} \in \mathbb{C} ^F$ at a time step $t$ for $c$-th channel as follows;
\begin{align}
Z = \mathrm{BLSTM}(\{ \bar{\mathbf{x}}_{t} \}_{t=1}^{T}), \label{eq:filter_blstm} \\
\Re(\mathbf{g}_{t,c}) = \mathrm{tanh}(\mathbf{W}_{c}^{\Re} \mathbf{z}_{t} + \mathbf{b}_{c}^{\Re}), \\
\Im(\mathbf{g}_{t,c}) = \mathrm{tanh}(\mathbf{W}_{c}^{\Im} \mathbf{z}_{t} + \mathbf{b}_{c}^{\Im}),
\end{align}
where $Z = \{\mathbf{z}_{t} \in \mathbb{R} ^{D_\text{Z}} | t=1, \cdots, T\}$　is a sequence of $D_\text{Z}$-dimensional output vectors of the BLSTM network.
$\bar{\mathbf{x}}_{t} = \{ \Re({x_{t,f,c}}), \Im({x_{t,f,c}}) \}_{f=1, c=1}^{F, C} \in \mathbb{R}^{2FC}$ is an input feature of a $2FC$-dimensional real-value vector for the BLSTM network.
This is obtained by concatenating the real and imaginary parts of all STFT coefficients in all channels.
$\Re(\mathbf{g}_{t,c})$ and $\Im(\mathbf{g}_{t,c})$ is the real and imaginary part of filter coefficients, $\mathbf{W}_{c}^{\Re} \in \mathbb{R}^{F \times D_{\text{Z}}}$ and $\mathbf{W}_{c}^{\Im} \in \mathbb{R}^{F \times D_{\text{Z}}}$ are the weight matrices of the output layer for $c$-th channel, and $\mathbf{b}_{c}^{\Re} \in \mathbb{R}^{F}$ and $\mathbf{b}_{c}^{\Im} \in \mathbb{R}^{F}$ are their corresponding bias vectors.
Using the estimated filters $\mathbf{g}_{t,c}$, the enhanced STFT coefficients $\hat{x}_{t,f}$ are obtained based on Eq.~(\ref{eq:enhance}).

This approach has several possible problems due to its formalization.
The first issue is the high flexibility of the estimated filters $\{g_{t,f,c}\}_{t=1,f=1,c=1}^{T,F,C}$, which are composed of a large number of unconstrained variables ($2TFC$) estimated from few observations.
This causes problems such as training difficulties and over-fitting.
The second issue is that the network structure depends on the number and order of channels.
Therefore, a new filter estimation network has to be trained when we change microphone configurations.

\subsection{Mask estimation network approach}
\label{sec:mask_net}

\begin{figure}[t]
\begin{center}
\centerline{\includegraphics[width=\columnwidth]{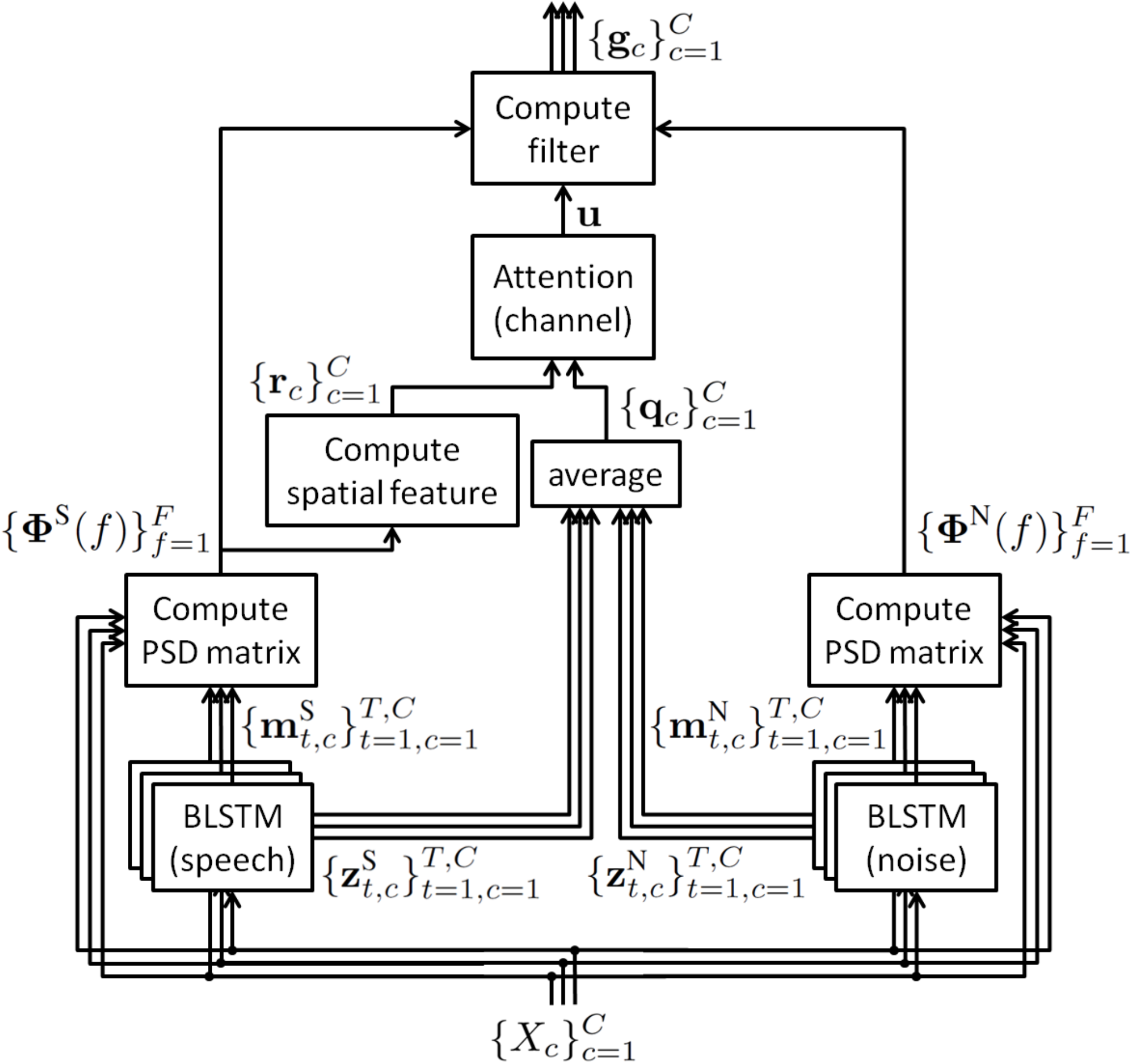}}
\caption{Overall procedures to compute filter coefficients in mask estimation network approach.}
\label{fig:mask_net_detail}
\end{center}
\vskip -0.2in
\end{figure} 

The key point of the mask estimation network approach is that it constrains the estimated filters based on well-founded  array signal processing principles.
Here, the network estimates the time-frequency masks, which are used to compute the \textit{time-invariant} filter coefficients $\{g_{f,c}\}_{f=1,c=1}^{F,C}$ based on the MVDR formalizations.
This is the main difference between this approach and the filter estimation network approach described in Section~\ref{sec:filter_net}.
Also, mask-based beamforming approaches have achieved great performance in noisy speech recognition benchmarks \cite{yoshioka2015ntt,heymann2016neural,erdogan2016improved}.
Therefore, this paper proposes to use a mask-based MVDR beamformer, where overall procedures are formalized as a differentiable network for the subsequent end-to-end speech recognition system.
Figure~\ref{fig:mask_net_detail} summarizes the overall procedures to compute the filter coefficients, which is a detailed flow of Figure \ref{fig:two_beamformer} (b).

\subsubsection{Mask-based MVDR formalization}
One of the MVDR formalizations computes the time-invariant filter coefficients $\mathbf{g}(f) = \{g_{f,c}\}_{c=1}^{C} \in \mathbb{C}^{C}$ in Eq.~(\ref{eq:enhance}) as follows \cite{souden2010optimal}:
\begin{align}
\mathbf{g}(f) = \frac{\boldsymbol{\Phi}^{\text{N}}(f)^{-1} \boldsymbol{\Phi}^{\text{S}}(f)}{\mathrm{Tr}(\boldsymbol{\Phi}^{\text{N}}(f)^{-1} \boldsymbol{\Phi}^{\text{S}}(f))} \mathbf{u},
\label{eq:mvdr}
\end{align}
where $\boldsymbol{\Phi}^{\text{S}}(f) \in \mathbb{C}^{C \times C}$ and $\boldsymbol{\Phi}^{\text{N}}(f) \in \mathbb{C}^{C \times C}$ are the cross-channel power spectral density (PSD) matrices (also known as spatial covariance matrices) for speech and noise signals, respectively. 
$\mathbf{u} \in \mathbb{R}^{C}$ is the one-hot vector representing a reference microphone, and $\mathrm{Tr}(\cdot)$ is the matrix trace operation.
Note that although the formula contains a matrix inverse, the number of channels is relatively small, and so the forward pass and derivatives can be efficiently computed.  

Based on \cite{yoshioka2015ntt,heymann2016neural}, the PSD matrices are robustly estimated using the expectation with respect to time-frequency masks as follows:
\begin{align}
\boldsymbol{\Phi}^{\text{S}}(f) = \frac{1}{\sum_{t=1}^{T} m_{t,f}^{\text{S}}} \sum_{t=1}^{T} m_{t,f}^{\text{S}} \mathbf{x}_{t,f} \mathbf{x}_{t,f}^{\dagger},
\label{eq:psd_speech}
\\
\boldsymbol{\Phi}^{\text{N}}(f) = \frac{1}{\sum_{t=1}^{T} m_{t,f}^{\text{N}}} \sum_{t=1}^{T} m_{t,f}^{\text{N}} \mathbf{x}_{t,f} \mathbf{x}_{t,f}^{\dagger},
\label{eq:psd_noise}
\end{align}
where $\mathbf{x}_{t,f} = \{x_{t,f,c}\}_{c=1}^{C} \in \mathbb{C}^{C}$ is the spatial vector of an observed signal for each time-frequency bin, $m_{t,f}^{\text{S}} \in [0, 1]$ and $m_{t,f}^{\text{N}} \in [0, 1]$ are the time-frequency masks for speech and noise, respectively. 
$\dagger$ represents the conjugate transpose.

\subsubsection{Mask estimation network}

In the mask estimation network approach, we use two real-valued BLSTM networks; one for a speech mask and the other for a noise mask.
Each network outputs the time-frequency mask as follows:
\begin{align}
Z_{c}^{\text{S}} &= \mathrm{BLSTM}^{\text{S}}(\{\bar{\mathbf{x}}_{t,c}\}_{t=1}^{T}), \label{eq:smask} \\
\mathbf{m}_{t,c}^{\text{S}} &= \mathrm{sigmoid}(\mathbf{W}^{\text{S}} \mathbf{z}_{t,c}^{\text{S}} + \mathbf{b}^{\text{S}}), \\
Z_{c}^{\text{N}} &= \mathrm{BLSTM}^{\text{N}}(\{\bar{\mathbf{x}}_{t,c}\}_{t=1}^{T}), \label{eq:nmask} \\
\mathbf{m}_{t,c}^{\text{N}} &= \mathrm{sigmoid}(\mathbf{W}^{\text{N}} \mathbf{z}_{t,c}^{\text{N}} + \mathbf{b}^{\text{N}}),
\end{align}
where $Z_{c}^{\text{S}} = \{\mathbf{z}_{t,c}^{\text{S}} \in \mathbb{R} ^{D_\text{Z}} | t=1, \cdots, T\}$ is the output sequence of $D_{\text{Z}}$-dimensional vectors of the BLSTM network to obtain a speech mask over $c$-th channel's input STFTs.
$Z_{c}^{\text{N}}$ is the BLSTM output sequence for a noise mask.
$\bar{\mathbf{x}}_{t,c} = \{ \Re({x_{t,f,c}}), \Im({x_{t,f,c}}) \}_{f=1}^{F} \in \mathbb{R}^{2F}$ is an input feature of a $2F$-dimensional real-value vector.
This is obtained by concatenating the real and imaginary parts of all STFT features at $c$-th channel.
$\mathbf{m}_{t,c}^{\text{S}} = \{m_{t,f,c}^{\text{S}}\}_{f=1}^{F} \in [0, 1]^{F}$ and $\mathbf{m}_{t,c}^{\text{N}}$ are the estimated speech and noise masks for every $c$-th channel at a time step $t$, respectively.
$\mathbf{W}^{\text{S}}, \mathbf{W}^{\text{N}} \in \mathbb{R}^{F \times D_{\text{Z}}}$ are the weight matrices of the output layers to finally output speech and noise masks, respectively, and $\mathbf{b}^{\text{S}}, \mathbf{b}^{\text{N}} \in \mathbb{R}^{F}$ are their corresponding bias vectors.

After computing the speech and noise masks for each channel, the averaged masks are obtained as follows:
\begin{align}
\mathbf{m}_{t}^{\text{S}} = \frac{1}{C} \sum_{c=1}^{C} \mathbf{m}_{t,c}^{\text{S}}, 
\quad 
\mathbf{m}_{t}^{\text{N}} = \frac{1}{C} \sum_{c=1}^{C} \mathbf{m}_{t,c}^{\text{N}}.
\end{align}
We use these averaged masks to estimate the PSD matrices as described in Eqs.~(\ref{eq:psd_speech}) and (\ref{eq:psd_noise}).
The MVDR beamformer through this BLSTM mask estimation is originally proposed in \cite{heymann2016neural}, but our neural beamformer further extends it with attention-based reference selection, which is described in the next subsection.

\subsubsection{Attention-based reference selection}
\label{sec:att_ref}
To incorporate the reference microphone selection in a neural beamformer framework, we use a soft-max for the vector $\mathbf{u}$ in Eq.~\eqref{eq:mvdr} derived from an attention mechanism.
In this approach, the reference microphone vector $\mathbf{u}$ is estimated from time-invariant feature vectors $\mathbf{q}_{c}$ and $\mathbf{r}_{c}$ as follows:
%
\begin{align}
\tilde{k}_{c} &= \mathbf{v}^{\mathrm{T}} \mathrm{tanh}(\mathbf{V}^{\text{Q}} \mathbf{q}_{c} + \mathbf{V}^{\text{R}}\mathbf{r}_{c} + \tilde{\mathbf{b}}), \\
u_{c} &= \frac{\exp(\beta \tilde{k}_{c})}{\sum_{c=1}^{C} \exp(\beta \tilde{k}_{c})}, \label{eq:att_ref_end}
\end{align}
where 
$\mathbf{v} \in \mathbb{R}^{1 \times D_{\text{V}}}, 
\mathbf{V}^{\text{Z}} \in \mathbb{R}^{D_{\text{V}} \times 2D_{\text{Z}}}, 
\mathbf{V}^{\text{R}} \in \mathbb{R}^{D_{\text{V}} \times 2F}$ 
are trainable weight parameters, 
$\tilde{\mathbf{b}} \in \mathbb{R}^{D_{\text{V}}}$ is a trainable bias vector.
$\beta$ is the sharpening factor.
We use two types of features; 1) the time-averaged state vector $\mathbf{q}_{c} \in \mathbf{R} ^{2D_{\text{Z}}}$ extracted from the BLSTM networks for speech and noise masks in Eqs.~\eqref{eq:smask} and \eqref{eq:nmask}, i.e., 
\begin{align}
 \mathbf{q}_{c} = \frac{1}{T} \sum_{t=1}^{T}\{ \mathbf{z}_{t,c}^{\text{S}}, \mathbf{z}_{t,c}^{\text{N}} \} \label{eq:att_ref_begin},
\end{align}
and 2) the PSD feature $\mathbf{r}_{c} \in \mathbb{R}^{2F}$, which incorporates the spatial information into the attention mechanism.
The following equation represents how to compute $\mathbf{r}_{c}$:
\begin{align}
\mathbf{r}_{c} = \frac{1}{C-1} \sum_{c'=1, c' \neq c}^{C} \{\Re(\phi_{f,c,c'}^{\text{S}}), \Im(\phi_{f,c,c'}^{\text{S}}) \}_{f=1}^{F},
\end{align}
where $\phi_{f,c,c'}^{S} \in \mathbb{C}$ is the entry in $c$-th row and $c'$-th column of the speech PSD matrix $\boldsymbol{\Phi}^{\text{S}}(f)$ in Eq.~(\ref{eq:psd_speech}).
The PSD matrix represents correlation information between channels.
To select a reference microphone, the spatial correlation related to speech signals is more informative, and therefore, we only use the speech PSD matrix $\boldsymbol{\Phi}^{\text{S}}(f)$ as a feature.

Note that, in this mask estimation based MVDR beamformer, masks for each channel are computed separately using the same BLSTM network unlike Eq.~(\ref{eq:filter_blstm}), and the mask estimation network is independent of channels.
Similarly, the reference selection network is also independent of channels, and the beamformer deals with input signals with arbitrary number and order of channels without re-training or re-configuration of the network.


\section{Multichannel end-to-end ASR}
\label{sec:me2e}

In this work, we propose a multichannel end-to-end speech recognition, which integrates all components with a single neural architecture.
We adopt neural beamformers (Section \ref{sec:beamforming_network}) as a speech enhancement part, and the attention-based encoder-decoder (Section \ref{sec:e2e_conventional}) as a speech recognition part.

The entire procedure to generate the sequence of output labels $\hat{Y}$ from the multichannel inputs $\{X_{c}\}_{c=1}^{C}$ is formalized as follows:
\begin{align}
\hat{X} &= \mathrm{Enhance}(\{X_{c}\}_{c=1}^{C}), \\
\hat{O} &= \mathrm{Feature}(\hat{X}), \\
\hat{H} &= \mathrm{Encoder}(\hat{O}), \\
\hat{\mathbf{c}}_{n} &= \mathrm{Attention}(\hat{\mathbf{a}}_{n-1}, \hat{\mathbf{s}}_{n}, \hat{H}), \\
\hat{y}_{n} &= \mathrm{Decoder}(\hat{\mathbf{c}}_{n}, \hat{y}_{1:n-1}).
\end{align}
$\mathrm{Enhance}(\cdot)$ is a speech enhancement function realized by the neural beamformer based on Eq.~\eqref{eq:enhance} with the filter or mask estimation network (Section \ref{sec:filter_net} or \ref{sec:mask_net}).

$\mathrm{Feature}(\cdot)$ is a feature extraction function.
In this work, we use a normalized log Mel filterbank transform to obtain $\hat{\mathbf{o}}_{t} \in \mathbb{R}^{D_{\text{O}}}$ computed from the enhanced STFT coefficients $\hat{\mathbf{x}}_{t} \in \mathbb{C} ^F$ as an input of attention-based encoder-decoder:
\begin{align}
\mathbf{p}_{t} &= \{\Re(\hat{x}_{t,f})^{2} + \Im(\hat{x}_{t,f})^{2}\}_{f=1}^{F}, \\
\hat{\mathbf{o}}_{t} &= \mathrm{Norm}(\log(\mathrm{Mel}(\mathbf{p}_{t}))),
\end{align}
where $\mathbf{p}_{t} \in \mathbb{R}^{F}$ is a real-valued vector of the power spectrum of the enhanced signal at a time step $t$, $\mathrm{Mel}(\cdot)$ is the operation of $D_{\text{O}} \times F$ Mel matrix multiplication, and $\mathrm{Norm}(\cdot)$ is the operation of global mean and variance normalization so that its mean and variance become 0 and 1.

$\mathrm{Encoder}(\cdot)$, $\mathrm{Attention}(\cdot)$, and $\mathrm{Decoder}(\cdot)$ are defined in Eqs.~\eqref{eq:enc}, \eqref{eq:att}, and \eqref{eq:dec}, respectively, with the sequence of the enhanced log Mel filterbank like features $\hat{O}$ as an input. 

Thus, we can build a multichannel end-to-end speech recognition system, which converts multichannel speech signals to texts with a single network.
Note that because all procedures, such as enhancement, feature extraction, encoder, attention, and decoder, are connected with differentiable graphs, we can optimize the overall inference to generate a correct label sequence.

\subsection*{Relation to prior works}
There have been several related studies of neural beamformers based on the filter estimation \cite{li2016neural,xiao2016deep} and the mask estimation \cite{heymann2016neural,erdogan2016improved,xiao2016study}.
The main difference is that such preceding studies use a component-level training objective within the conventional hybrid frameworks, while our work focuses on the entire end-to-end objective.
For example, \citealp{heymann2016neural,erdogan2016improved} use a signal-level objective (binary mask classification or regression) to train a network given parallel clean and noisy speech data.
\citealp{li2016neural,xiao2016deep,xiao2016study} use ASR objectives (HMM state classification or sequence discriminative training), but they are still based on the hybrid approach.
Speech recognition with raw multichannel waveforms \cite{hoshen2015speech,sainath2016reducing} can also be seen as using a neural beamformer, where the filter coefficients are represented as network parameters, but again these methods are still based on the hybrid approach.

As regards end-to-end speech recognition, all existing studies are based on a single channel setup.
For example, most studies focus on a standard clean speech recognition setup without speech enhancement. \cite{chorowski2014end,graves2014towards,chorowski2015attention,chan2016listen,miao2015eesen,zhang2016very,kim2016joint,lu2016training}.
\citealp{amodei2016deep} discusses end-to-end speech recognition in a noisy environment, but this method deals with the noise robustness by preparing various types of simulated noisy speech for training data, and does not incorporate multichannel speech enhancement in their networks.

\section{Experiments}
We study the effectiveness of our multichannel end-to-end system compared to  a baseline end-to-end system with noisy speech or beamformed inputs.
We use the two multichannel speech recognition benchmarks, CHiME-4 \cite{vincent2016analysis} and AMI \cite{hain2007ami}.

CHiME-4 is a speech recognition task in public noisy environments, consisting of speech recorded using a tablet device with 6-channel microphones.
It consists of real and simulated data.
The training set consists of 3 hours of real speech data uttered by 4 speakers and 15 hours of simulation speech data uttered by 83 speakers.
The development set consists of 2.9 hours of real and simulation speech data uttered by 4 speakers, respectively.
The evaluation set consists of 2.2 hours of real and simulation speech data uttered by 4 speakers, respectively.
We excluded the 2nd channel signals, which is captured at the microphone located on the backside of the tablet, and used 5 channels for the following multichannel experiments ($C=5$).

AMI is a speech recognition task in meetings, consisting of speech recorded using 8-channel circular microphones ($C=8$).
It consists of only real data.
The training set consists of about 78 hours of speech data uttered by 135 speakers.
the development and evaluation sets consist of about 9 hours of speech data uttered by 18 and 16 speakers, respectively.
The amount of training data (i.e., 78 hours) is larger than one for CHiME-4 (i.e., 18 hours), and we mainly used CHiME-4 data to demonstrate our experiments.

\subsection{Configuration}

\subsubsection{Encoder-decoder networks}
We used 40-dimensional log Mel filterbank coefficients as an input feature vector for both noisy and enhanced speech signals ($D_{\text{O}} = 40$).
In this experiment, we used $4$-layer BLSTM with $320$ cells in the encoder ($D_{\text{H}} = 320$), and $1$-layer LSTM with $320$ cells in the decoder ($D_{\text{S}} = 320$).
In the encoder, we subsampled the hidden states of the first and second layers and used every second of hidden states for the subsequent layer's inputs.
Therefore, the number of hidden states at the encoder's output layer is reduced to $L=T/4$.
After every BLSTM layer, we used a linear projection layer with $320$ units to combine the forward and backward LSTM outputs.
For the attention mechanism, $10$ centered convolution filters ($D_{\text{F}} = 10$) of width $100$ ($D_{\text{f}} = 100$) were used to extract the convolutional features.
We set the attention inner product dimension as 320 ($D_{\text{W}}=320$), and used the sharpening factor $\alpha = 2$.
To boost the optimization in a noisy environment, we adopted a joint CTC-attention multi-task loss function \cite{kim2016joint}, and set the CTC loss weight as $0.1$.

For decoding, we used a beam search algorithm similar to \cite{sutskever2014sequence} with the beam size $20$ at each output step to reduce the computation cost.
CTC scores were also used to re-score the hypotheses with $0.1$ weight.
We adopted a length penalty term \cite{chorowski2015attention} to the decoding objective and set the penalty weight as $0.3$.
In the CHiME-4 experiments, we only allowed the hypotheses whose length were within $0.3 \times L$ and $0.75 \times L$ during decoding, while the hypothesis lengths in the AMI experiments were automatically determined based on the above scores.
Note that we pursued a pure end-to-end setup without using any external lexicon or language models, and used CER as an evaluation metric.

\subsubsection{Neural beamformers}
256 STFT coefficients and the offset were computed from 25ms-width hamming window with 10ms shift ($F = 257$).
Both filter and mask estimation network approaches used similar a $3$-layer BLSTM with $320$ cells ($D_{\text{Z}} = 320$) without the subsampling technique.
For the reference selection attention mechanism, we used the same attention inner product dimension ($D_{\text{V}} = 320$) and sharpening factor $\beta = 2$ as those of the encoder-decoder network.

\subsubsection{Shared configurations}

All the parameters are initialized with the range [-0.1, 0.1] of a uniform distribution.
We used the AdaDelta algorithm \cite{zeiler2012adadelta} with gradient clipping \cite{pascanu2013difficulty} for optimization.
We initialized the AdaDelta hyperparameters $\rho = 0.95$ and $\epsilon = 1^{-8}$.
Once the loss over the validation set was degraded, we decreased the AdaDelta hyperparameter $\epsilon$ by multiplying it by $0.01$ at each subsequent epoch.
The training procedure was stopped after 15 epochs.
During the training, we adopted multi-condition training strategy, i.e., in addition to the optimization with the enhanced features through the neural beamformers, we also used the noisy multichannel speech data as an input of encoder-decoder networks without through the neural beamformers to improve the robustness of the encoder-decoder networks.
All the above networks are implemented by using Chainer \cite{tokui2015chainer}.

\subsection{Results}

\begin{table}[t]
\caption{Character error rate [\%] for CHiME-4 corpus.}
\label{table:chime4}
\vskip 0.15in
\begin{center}
\begin{small}
\begin{sc}
\scalebox{1.0}[1.0]{
\begin{tabular}{c|cccc}
\hline
\abovespace\belowspace
model & \raisebox{-1mm}{\shortstack{dev \\simu}} & \raisebox{-1mm}{\shortstack{dev \\real}} & \raisebox{-1mm}{\shortstack{eval \\simu}} & \raisebox{-1mm}{\shortstack{eval \\real}} \\
\hline
\abovespace
noisy & 25.0 & 24.5 & 34.7 & 35.8 \\
\belowspace
beamformit & 21.5 & 19.3 & 31.2 & 28.2 \\
\hline
\abovespace
filter\_net & 19.1 & 20.3 & 28.2 & 32.7 \\
mask\_net (ref) & 15.5 & 18.6 & \bf{23.7} & 28.8 \\
\belowspace
mask\_net (att) & \bf{15.3} & \bf{18.2} & \bf{23.7} & \bf{26.8} \\
\hline
\end{tabular}
}
\end{sc}
\end{small}
\end{center}
\vskip -0.2in
\end{table}

Table~\ref{table:chime4} shows the recognition performances of CHiME-4 with the five systems: NOISY, BEAMFORMIT, FILTER\_NET, MASK\_NET (REF), and MASK\_NET (ATT).
NOISY and BEAMFORMIT were the baseline single-channel end-to-end systems, which did not include the speech enhancement part in their frameworks.
Their end-to-end networks were trained only with noisy speech data by following a conventional multi-condition training strategy \cite{vincent2016analysis}.
During decoding, NOISY used single-channel noisy speech data from 'isolated 1ch track' in CHiME-4 as an input, while BEAMFORMIT used the enhanced speech data obtained from 5-channel signals with BeamformIt \cite{anguera2007acoustic}, which is well-known delay-and-sum beamformer, as an input.

FILTER\_NET, MASK\_NET (REF), and MASK\_NET (ATT) were the multichannel end-to-end systems described in Section~\ref{sec:me2e}.
To evaluate the validity of the reference selection, we prepared MASK\_NET (ATT) based on the mask-based beamformer with attention-based reference selection described in Section~\ref{sec:att_ref}, and MASK\_NET (REF) with $5$-th channel as a fixed reference microphone, which is located on the center front of the tablet device.

Table~\ref{table:chime4} shows that BEAMFORMIT, FILTER\_NET, MASK\_NET (REF), and MASK\_NET (ATT) outperformed NOISY, which confirms the effectiveness of combining speech enhancement with the attention-based encoder-decoder framework.
The comparison of MASK\_NET (REF) and MASK\_NET (ATT) validates the use of the attention-based mechanism for reference selection.
FILTER\_NET, which is based on the filter estimation network described in Section \ref{sec:filter_net}, also improved the performance compared to NOISY, but worse than MASK\_NET (ATT).
This is because it is difficult to optimize the filter estimation network due to a lack of restriction to estimate filter coefficients, and it needs some careful optimization, as suggested by \cite{xiao2016deep}.
Finally, MASK\_NET (ATT) achieved better recognition performance than BEAMFORMIT, which proves the effectiveness of our joint integration rather than a pipe-line combination of speech enhancement and (end-to-end) speech recognition.

\begin{table}[t]
\caption{Character error rate [\%] for AMI corpus.}
\label{table:ami}
\vskip 0.15in
\begin{center}
\begin{small}
\begin{sc}
\scalebox{1.0}[1.0]{
\begin{tabular}{c|cc}
\hline
\abovespace\belowspace
model & dev & eval \\
\hline
\abovespace
noisy & 41.8 & 45.3 \\
\belowspace
beamformit & 44.9 & 51.3 \\
\hline
\abovespace\belowspace
mask\_net (att) & \bf{35.7} & \bf{39.0} \\
\hline
\end{tabular}
}
\end{sc}
\end{small}
\end{center}
\vskip -0.2in
\end{table}

To further investigate the effectiveness of our proposed multichannel end-to-end framework, we also conducted the experiment on the AMI corpus.
Table~\ref{table:ami} compares the recognition performance of the three systems: NOISY, BEAMFORMIT, and MASK\_NET (ATT).
In NOISY, we used noisy speech data from the 1st channel in AMI as an input to the system.
Table~\ref{table:ami} shows that, even in the AMI, our proposed MASK\_NET (ATT) achieved better recognition performance than the attention-based baselines (NOISY and BEAMFORMIT), which also confirms the effectiveness of our proposed multichannel end-to-end framework.
Note that BEAMFORMIT was worse than NOISY even with the enhanced signals.
This phenomenon is sometimes observed in noisy speech recognition that the distortion caused by sole speech enhancement degrades the performance without re-training.
Our end-to-end system jointly optimizes the speech enhancement part with the ASR objective, and can avoid such degradations.

\subsection{Influence on the number and order of channels}
As we discussed in Section \ref{sec:mask_net}, one unique characteristic of our proposed MASK\_NET (ATT) is the robustness/invariance against the number and order of channels without re-training.
Table~\ref{table:channels} shows an influence of the CHiME-4 validation accuracies on the number and order of channels.
The validation accuracy was computed conditioned on the ground truth labels $y_{1:n-1}^{*}$ in Eq.~\eqref{eq:e2eobj} during decoder's recursive character generation, which has a strong correlation with CER.
The second column of the table represents the channel indices, which were used as an input of the same MASK\_NET (ATT) network.

Comparison of $5\_6\_4\_3\_1$ and $3\_4\_1\_5\_6$ shows that the order of channels did not affect the recognition performance of MASK\_NET (ATT) at all, as we expected.
In addition, even when we used fewer three or four channels as an input, MASK\_NET (ATT) still outperformed NOISY (single channel).
These results confirm that our proposed multichannel end-to-end system can deal with input signals with arbitrary number and order of channels, without any re-configuration and re-training.

\begin{table}[t]
\caption{CHiME-4 validation accuracies [\%] for MASK\_NET (ATT) with different numbers and orders of channels.}
\label{table:channels}
\begin{center}
\begin{small}
\begin{sc}
\scalebox{1.0}[1.0]{
\begin{tabular}{c|c|c}
\hline
\abovespace\belowspace
model & channel & dev \\
\hline
\abovespace\belowspace
noisy & isolated\_1ch\_track & 87.9 \\
\hline
\abovespace
mask\_net (att) & 5\_6\_4\_3\_1 & 91.2 \\
\belowspace
mask\_net (att) & 3\_4\_1\_5\_6 & 91.2 \\
\hline
\abovespace
mask\_net (att) & 5\_6\_4\_1 & 91.1 \\
\belowspace
mask\_net (att) & 5\_6\_4 & 90.9 \\
\hline
\end{tabular}
}
\end{sc}
\end{small}
\end{center}
\vskip -0.2in
\end{table}

\subsection{Visualization of beamformed features}
\begin{figure}[t]
\begin{center}
\centerline{\includegraphics[width=.9\columnwidth]{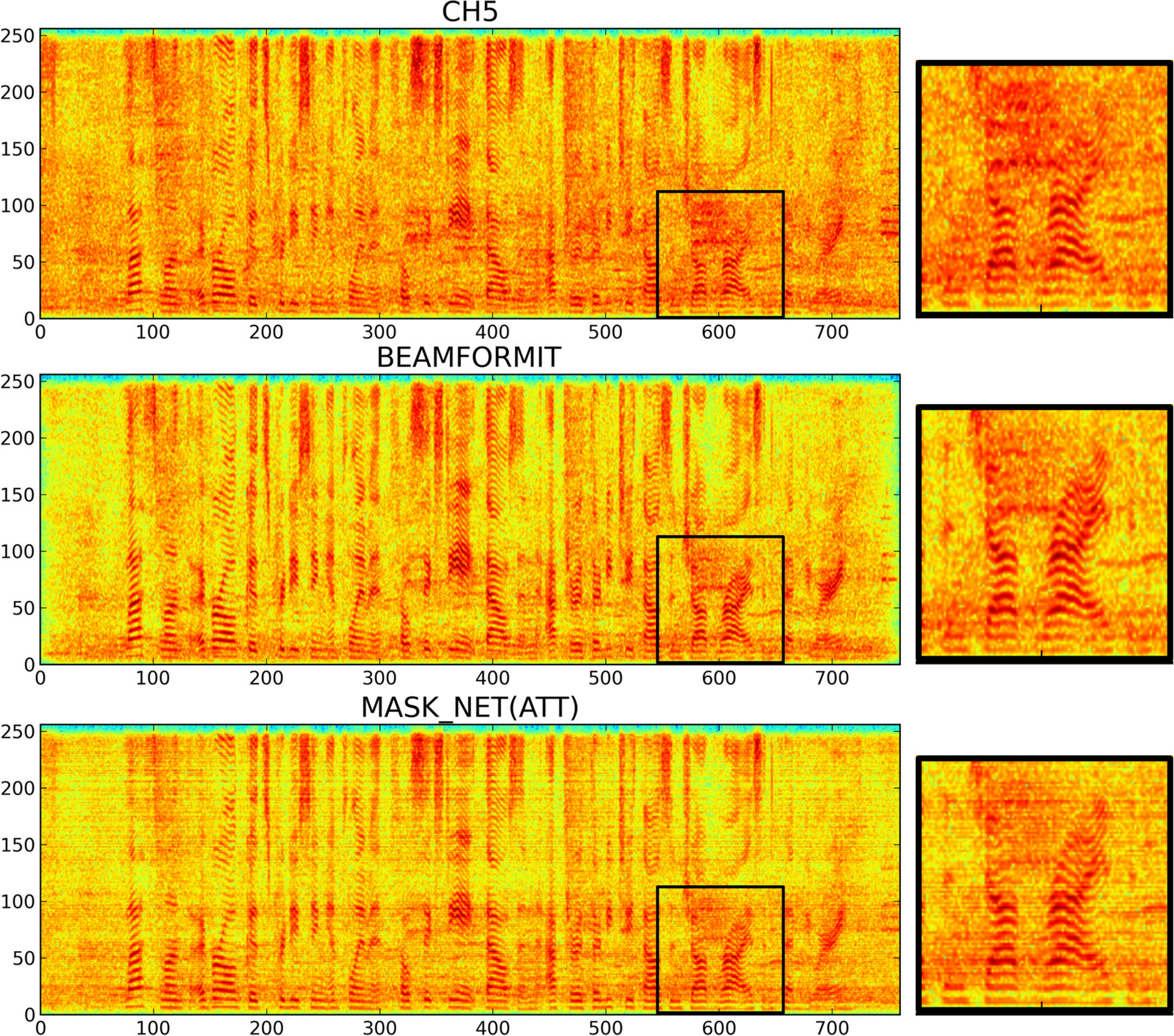}}
\caption{Comparison of the log-magnitude spectrograms of the same CHiME-4 utterance with the 5-th channel noisy signal, enhanced signal with BeamformIt, and enhanced signal with our proposed MASK\_NET (ATT).}
\label{fig:spectrogram}
\end{center}
\vskip -0.2in
\end{figure}

To analyze the behavior of our developed speech enhancement component with a neural beamformer, Figure~\ref{fig:spectrogram} visualizes the spectrograms of the same CHiME-4 utterance with the 5-th channel noisy signal, enhanced signal with BeamformIt, and enhanced signal with our proposed MASK\_NET (ATT).
We could confirm that the BeamformIt and MASK\_NET (ATT) successfully suppressed the noises comparing to the 5-th channel signal by eliminating blurred red areas overall. 
In addition, by focusing on the insides of black boxes, the harmonic structure, which was corrupted in the 5-th channel signal, was recovered in BeamformIt and MASK\_NET (ATT).

This result suggests that our proposed MASK\_NET (ATT) successfully learned a noise suppression function similar to the conventional beamformer, although it is optimized based on the end-to-end ASR objective, without explicitly using clean data as a target.

\section{Conclusions}

In this paper, we extended an existing attention-based encoder-decoder framework by integrating a neural beamformer and proposed a multichannel end-to-end speech recognition framework.
It can jointly optimize the overall inference in multichannel speech recognition (i.e., from speech enhancement to speech recognition) based on the end-to-end ASR objective, and it can generalize to different numbers and configurations of microphones. 
The experimental results on challenging noisy speech recognition benchmarks, CHiME-4 and AMI, show that the proposed framework outperformed the end-to-end baseline with noisy and delay-and-sum beamformed inputs.

The current system still has data sparseness issues due to the lack of lexicon and language models, unlike the conventional hybrid approach.
Therefore, the results reported in the paper did not reach the state-of-the-art performance in these benchmarks, but they are still convincing to show the effectiveness of the proposed framework. 
Our most important future work is to overcome these data sparseness issues by developing adaptation techniques of an end-to-end framework with the incorporation of linguistic resources.

 

\clearpage

\bibliography{icml}
\bibliographystyle{icml2017}

\end{document}